\documentclass[a4paper]{article}

\usepackage{INTERSPEECH2021}
\usepackage{bm}
\usepackage{graphicx}
\usepackage{float}
\usepackage{ascmac} 
\usepackage{fancybox}
\usepackage{float}
\usepackage{tabularx}
\usepackage{booktabs}
\usepackage{amssymb}
\usepackage{multirow}
\usepackage{cite}
\usepackage{multirow}
\usepackage[table,xcdraw]{xcolor}
\usepackage[normalem]{ulem}
\useunder{\uline}{\ul}{}
\usepackage{url}
\usepackage{pifont}

\newcommand{\xmark}{\ding{55}}

\AtBeginDocument{
  \abovedisplayskip     =.8\abovedisplayskip
  \abovedisplayshortskip=.8\abovedisplayshortskip
  \belowdisplayskip     =.8\belowdisplayskip
  \belowdisplayshortskip=.8\belowdisplayshortskip}

\title{Streaming Target-Speaker ASR with Neural Transducer}
\name{Takafumi Moriya$^{1,2 \ast}$ \thanks{$^{\ast}$ Equal contribution.}, Hiroshi Sato$^{1 \ast}$, Tsubasa Ochiai$^{1}$, Marc Delcroix$^{1}$, Takahiro Shinozaki$^{2}$}
\address{\begin{tabular}{c} $^{1}$NTT Corporation, Japan; $^{2}$Tokyo Institute of Technology, Japan \end{tabular}}
\email{\{takafumi.moriya.nd,hiroshi.satou.bh\}@hco.ntt.co.jp}

\ninept

\begin{document}

\maketitle
\begin{abstract} 
Although recent advances in deep learning technology have boosted automatic speech recognition (ASR) performance in the single-talker case, 
it remains difficult to recognize multi-talker speech in which many voices overlap. 
One conventional approach to tackle this problem is to use a cascade of a speech separation or target speech extraction front-end with an ASR back-end. 
However, the extra computation costs of the front-end module are a critical barrier to quick response, especially for streaming ASR. 
In this paper, we propose a target-speaker ASR (TS-ASR) system that implicitly integrates the target speech extraction functionality within a streaming end-to-end (E2E) ASR system, i.e. recurrent neural network-transducer (RNNT). 
Our system uses a similar idea as adopted for target speech extraction, but implements it directly at the level of the encoder of RNNT. 
This allows TS-ASR to be realized without placing extra computation costs on the front-end. 
Note that this study presents two major differences between prior studies on E2E TS-ASR; 
we investigate streaming models and base our study on Conformer models, 
whereas prior studies used RNN-based systems and considered only offline processing. 
We confirm in experiments that our TS-ASR achieves comparable recognition performance with conventional cascade systems in the offline setting, while reducing computation costs and realizing streaming TS-ASR. 
\end{abstract}

\section{Introduction}
End-to-end automatic speech recognition (E2E ASR), which can directly map acoustic features to output tokens, has recently attracted increasing interest from the ASR research community. 
Recent E2E ASR systems have achieved high recognition performance for single speaker conditions~\cite{anmol2020conformer,Sainath2021AnES,kurata2020kdrnnt,googleonline2020,chen2021lcconformer,moriya2021simple,moriya2021rnnts2s,moriya2022rnntadlm}. 
However, it remains challenging to recognize speech under severe conditions such as in presence of overlapping speakers~\cite{barker2018fifth}. 
This paper addresses the challenging problem of recognizing a target speaker in a mixture, i.e., target-speaker ASR (TS-ASR), in a streaming manner to support applications such as voice search or user-dependent voice interactive systems.

We can tackle this problem from the viewpoint of separation~\cite{hershey2016deep,luo2018tasnet,qian2018single,luo2019conv,li2014overview} or target speech extraction (TSE)~\cite{vzmolikova2019speakerbeam,delcroix2018single,wangvoicefilter,sato2021SE,Sato2021switch,ShiZW2021jointmtasr}, using a cascade speech enhancement front-end/ASR back-end system or an E2E system. 
Speech separation separates a signal into all of its sources~\cite{comon2010handbook,wang2018supervised}. 
It allows ASR of all speakers in a mixture. 
However, the problem of global permutation ambiguity means that the mapping between output and speaker is arbitrary. 
Moreover, decoding the speech of all speakers can be computationally involving. 
In many applications such as voice search, we are interested in recognizing the speech of a specific user and not the interfering speakers. 
In this case, TSE is a particularly attractive solution. 
TSE extracts only the speech of a target speaker given an enrollment of that speaker’s voice. 
It naturally avoids the output-speaker ambiguity and can limit the decoding to just the target speaker. 

Both separation and TSE approaches can be implemented as a cascade or within an E2E system~\cite{qian2018single,delcroix2018single,kanda2019auxiliary,denisov2019end,delcroix2019end,wangvoicefilter,sato2021SE,Sato2021switch,ShiZW2021jointmtasr}.
Cascade systems use a separation or TSE front-end that estimates the speech signals and then performs recognition. 
Cascade systems are modular, making it easy to visualize the different processing steps. 
However, although they can achieve high performance in offline modes, this comes at the expense of a high computational cost. 
Morever, the performance of separation and TSE front-ends degrade severely in streaming mode~\cite{luo2019conv,wangvoicefilter}. 
In contrast, E2E approaches directly process the mixture and output the recognition results without explicitly estimating the speech signals. 
They can achieve a similar level of performance at lower computational cost. 

Many ASR applications require fast and accurate responses in real-time. 
Recently, there has been increased interest in developing streaming E2E ASR systems. 
Recurrent neural network-transducer (RNNT)~\cite{Graves2012} is a promising technology for streaming E2E ASR applications, and has been extensively investigated for single speaker ASR~\cite{Sainath2021AnES,kurata2020kdrnnt,googleonline2020,chen2021lcconformer,moriya2021simple,moriya2021rnnts2s,moriya2022rnntadlm}. 
Several works have extended the ideas of RNNT for multi-talker ASR~\cite{SklyarPL21PIT,Lu2021surt,Kanda2022mtasr}. 
These works are E2E extensions of separation-based approaches. 
They recognize the speech of each speaker in a mixture, which is computationally demanding~\cite{SklyarPL21PIT,Lu2021surt}. 
Moreover, they do not identify the speakers and thus cannot be used directly for TS-ASR~\cite{SklyarPL21PIT,Lu2021surt,Kanda2022mtasr}. 
To the best of our knowledge, no study has investigated streaming E2E TS-ASR.

In this work, we propose an E2E TS-ASR framework that integrates the TSE functionality within an RNNT. 
We call this framework target-speaker RNNT (TS-RNNT). 
The system operates as follows. 
First, we extract a target speaker embedding from the enrollment using a speaker encoder module. 
The speaker embedding is used as an auxiliary input to the encoder of RNNT to inform the system which speaker to recognize in the mixture. 
We can incorporate the embedding within the RNNT encoder with a simple element-wise operation at an intermediate layer. 
Note that the embedding can be computed in advance. 
Therefore, our proposed TS-RNNT does not impact the computational complexity nor the latency of a vanilla RNNT model. 
This study presents two major differences from prior studies about E2E TS-ASR~\cite{denisov2019end,delcroix2019end}; we investigate streaming models and base our study on the recently proposed state-of-the-art Conformer models~\cite{anmol2020conformer}, whereas prior studies used long short-term memory (LSTM)-based systems and dealt only with offline processing.

We conduct experiments to compare our proposed TS-RNNT with a cascade combination of TSE and RNNT (TSE+RNNT) for offline and streaming modes. 
Experiments show that our proposed TS-RNNT achieves competitive performance with TSE+RNNT for offline conditions at significantly lower computational cost. 
More importantly, it greatly outperforms a TSE+RNNT system in streaming mode. 

\begin{figure}[t]
\vspace{-0.2cm}
 \begin{center}
    \includegraphics[width=8.1cm]{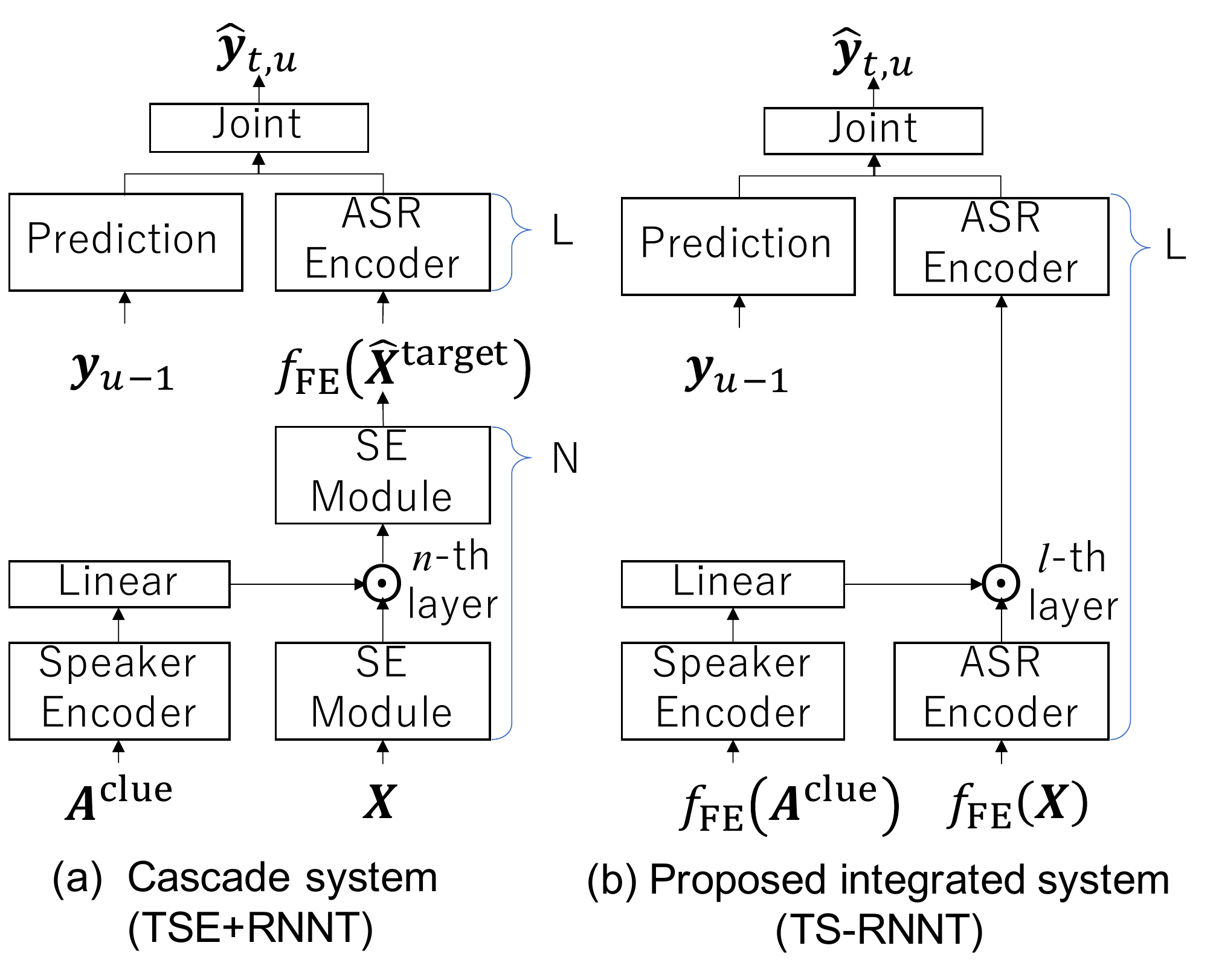}
 \end{center}
\vspace{-0.8cm}
 \caption{The overview of TS-ASR systems.}
 \vspace{-0.5cm}
 \label{fig:framework}
\end{figure}

\section{Target-Speaker ASR}
In this section, we introduce the TS-ASR systems used in our experiments. 
We first explain the TSE front-end and RNNT-based ASR back-end modules, which are the foundations of this work. Then we introduce our proposed TS-RNNT system.

Let $\bm{X} = \left[ x_1, ..., x_{T^{\prime}} \right] \in \mathcal{R}^{T^{\prime}}$ be the single microphone input speech mixture of duration $T^{\prime}$ that includes the target and interfering speaker's voices and noise. 
$Y = \left[ y_1, ..., y_U \right]$ is the sequence of tokens of length $U$ associated with the utterance spoken by target speaker, where $y_u \in \{1, ..., K\}$.
$K$ is the number of tokens. 

\vspace{-0.2cm}
\subsection{Overview of cascade TS-ASR (TSE+RNNT)}
\label{ssec:cascade}
\vspace{-0.1cm}
Fig.~\ref{fig:framework}~(a) is a diagram of the cascade TS-ASR system, which is composed of the TSE and ASR modules described below.

\vspace{-0.25cm}
\subsubsection{TSE front-end}
\label{sssec:se}
\vspace{-0.1cm}
The TSE module extracts the target-speaker's speech $\bm{X}^{\text{target}}$ from the mixture $\bm{X}$ using an enrollment speech $\bm{A}^{\text{clue}}$. We use the time-domain TSE system we developed in~\cite{sato2021SE,Sato2021switch}, which is similar to \cite{xu2019time,delcroix2020improving}.

First, we extract the embedding of the target-speaker, $\bm{h}^{\text{target}}$, from the enrollment speech, $\bm{A}^{\text{clue}}$, using a speaker encoder $f^{\text{Spk-Enc}}(\cdot)$ which consists of a multi-layer neural network followed by a linear layer and a time-average pooling layer. 
Then, we use an speaker extraction network, $f^{\text{SE}}(\cdot)$ to extract the target speech given the embedding, $\bm{h}^{\text{target}}$. 
The above operations, which yield the estimated target speech signal $\hat{\bm{X}}^{\text{target}}$, are defined as follows:
\begin{eqnarray}
\bm{H}^{\text{target}} &=& f^{\text{Spk-Enc}} (\bm{A}^{\text{clue}}; \theta^{\text{Spk-Enc}}), \\
\bm{h}^{\text{target}} &=& \frac{1}{T^{\prime}} \sum_{t^{\prime}=1}^{T^{\prime}} \bm{h}^{\text{target}}_{t^{\prime}}, \\
\hat{\bm{X}}^{\text{target}} &=& f^{\text{SE}} (\bm{X}, \bm{h}^{\text{target}}; \theta^{\text{SE}}). 
\label{eq:se}
\end{eqnarray}
In this work, we insert the embedding $\bm{h}^{\text{target}}$ at the $n$-th layer of $f^{\text{SE}} (\cdot)$ by performing the Hadamard product between the embedding and the output of that layer, $\bm{h}^{(n)}_{t^{\prime}}$, i.e. $\bm{h}^{(n)}_{t^{\prime}} \odot \bm{h}^{\text{target}}$. 
The learnable parameters $\theta^{\text{TSE}} \triangleq [\theta^{\text{Spk-Enc}}, \theta^{\text{SE}}]$ are jointly optimized by using the source-to-distortion ratio (SDR) loss~\cite{luo2019conv}.

\begin{table*}[th]
\vspace{-6pt}
\centering
\caption{Data generation setup. The number of utterances in (a) is double of (b) due to single-speaker case of (b).}
 \vspace{-0.25cm}
\label{tab:setup}
\scalebox{0.91}[0.91]{
\begin{tabular}{lllcccc}
\hline
 & \multicolumn{1}{c}{dataset} & \multicolumn{1}{c}{mixture type} & \begin{tabular}[c]{@{}c@{}}SIR\\ {[}dB{]}\end{tabular} & \begin{tabular}[c]{@{}c@{}}SNR\\ {[}dB{]}\end{tabular} & \begin{tabular}[c]{@{}c@{}}\#speakers\\ (train set / dev set)\end{tabular} & \begin{tabular}[c]{@{}c@{}}\#mixtures or utterances\\ (train set / dev set)\end{tabular} \\ \hline
(a) & training data for ASR backend (RNNT) & 1 speaker and noise & - & 0 - 20 & 3054 / 160 & 400,000 / 10,000 \\
(b) & training data for TSE and TS-RNNT & 2 speakers and noise & -5 - 5 & 0 - 20 & 3054 / 160 & 200,000 / 5,000 \\
(c) & evaluation data & 2 speakers and noise & -5 - 5 & 0, 5, 10, 15, 20 & 30 & $6,000 \times 5 = 30,000$ \\ \hline
\end{tabular}
}
\vspace{-0.4cm}
\end{table*}

\vspace{-0.15cm}
\subsubsection{ASR back-end}
\label{sssec:rnnt}
\vspace{-0.1cm}
We adopt an RNNT-based ASR back-end~\cite{Graves2012} that can perform streaming ASR. 
RNNT learns the mapping between sequences of different lengths. 
It consists of an ASR encoder and a prediction network, which allows the posterior probabilities to be jointly conditioned on not only the ASR encoder outputs but also on previous predictions. 
The processing is described as follows. 

First, the target speech signal, $\bm{X}^{\text{target}}$, is transformed into a sequence of acoustic features, i.e. log Mel-filterbank, using a feature extractor $f^{\text{FE}}(\cdot)$. 
Then, the features are encoded into length-$T$ sequence, $\bm{H}^{\text{ASR}} = \left[ \bm{h}^{\text{ASR}}_{1}, ..., \bm{h}^{\text{ASR}}_{T} \right]$, via an ASR encoder network $f^{\text{ASR-Enc}}(\cdot)$.
Next, the tokens, $Y$, are also encoded into
$\bm{H}^{\text{Pred}} = \left[ \bm{h}^{\text{Pred}}_{1}, ..., \bm{h}^{\text{Pred}}_{U} \right]$
via a prediction network $f^{\text{Pred}}(\cdot)$.
These two encoded features are fed to a feed-forward network, $f^{\text{Joint}}(\cdot)$, to compute the token posterior probabilities, $\hat{\bm{y}}_{t,u}$.
The above operation can be expressed as follows:
\begin{eqnarray}
\bm{h}^{\text{ASR}}_{t} &=& f^{\text{ASR-Enc}} (f^{\text{FE}}(x_{t^{\prime}}^{\text{target}}); \theta^{\text{ASR-Enc}}), \\
\bm{h}^{\text{Pred}}_{u} &=& f^{\text{Pred}} (y_{u-1}; \theta^{\text{Pred}}), \\
\hat{\bm{y}}_{t,u} &=& \text{Softmax} \left(f^{\text{Joint}} (\bm{h}^{\text{ASR}}_{t}, \bm{h}^{\text{Pred}}_{u}; \theta^{\text{Joint}}) \right),
\label{eq:rnnt}
\end{eqnarray}
where $\text{Softmax}(\cdot)$ indicates a softmax operation. 
All the learnable parameters $\theta^{\text{RNNT}} \triangleq [\theta^{\text{ASR-Enc}}, \theta^{\text{Pred}}, \theta^{\text{Joint}}]$ are optimized by using RNNT loss with the forward-backward algorithm~\cite{Graves2012}.

For training, we use the target speech signals, $\bm{X}^{\text{target}}$ corrupted with noise to make a robust system. 
At test time, we use the extracted signal estimated with TSE, $\hat{\bm{X}}^{\text{target}}$. 
Note that we do not optimize the TSE and ASR modules jointly. 
Joint optimization of TSE front-end and ASR for streaming applications will be part of our future works. 

\vspace{-0.1cm}
\subsection{Overview of integrated TS-ASR with RNNT}
\vspace{-0.1cm}
In this paper, we propose an integrated modeling approach for TS-ASR using RNNT, which operates in a fully E2E manner, and allows streaming ASR. 

\vspace{-0.1cm}
\subsubsection{Proposed TS-RNNT architecture}
\vspace{-0.1cm}
Fig.~\ref{fig:framework} (b) is a schematic diagram of our proposed TS-RNNT system. 
The architecture is almost the same as that of the vanilla RNNT described in~\ref{sssec:rnnt}. 
The difference is that the encoder of the TS-RNNT inputs directly the speech mixture and uses the target speaker embedding to inform which speaker in the mixture to decode. 
This is performed using a similar processing as the TSE front-end but within the encoder, i.e., a similar speaker encoder module and a fusion mechanism at an intermediate layer using the Hadamard product.

TS-RNNT encoder, $f^{\text{ASR-Enc}^{\prime}}(\cdot)$, is a modified version of ASR encoder $f^{\text{ASR-Enc}} (\cdot)$ that receives the speaker embedding extracted from speaker encoder $f^{\text{Spk-Enc}} (\cdot)$. 
$f^{\text{ASR-Enc}^{\prime}}(\cdot)$ is defined as follows:
\begin{eqnarray}
\bm{H}^{\text{target}^{\prime}} &=& f^{\text{Spk-Enc}^{\prime}} (f^{\text{FE}} (\bm{A}^{\text{clue}}); \theta^{\text{Spk-Enc}^{\prime}}), \\
\bm{h}^{\text{target}^{\prime}} &=& \frac{1}{T} \sum_{t=1}^{T} \bm{h}^{\text{target}^{\prime}}_{t}, \\
\bm{h}^{\text{ASR}^{\prime}}_{t} &=& f^{\text{ASR-Enc}^{\prime}} (f^{\text{FE}}(x_{t^{\prime}}), \bm{h}^{\text{target}^{\prime}}; \theta^{\text{ASR-Enc}^{\prime}} ), 
\label{eq:tsrnnt}
\end{eqnarray}
where $\bm{H}^{\text{target}^{\prime}}$ with length-$T$ is the speaker encoder outputs given the input $\bm{A}^{\text{clue}}$ as the enrollment, 
which is averaged with time axis and embedded into $\bm{h}^{\text{target}^{\prime}}$. 
$\bm{h}^{\text{target}^{\prime}}$ and $l$-th layer intermediate output $\bm{h}^{(l)}_{t}$ of the ASR encoder are multiplied as a Hadamard product, i.e. $\bm{h}^{(l)}_{t} \odot \bm{h}^{\text{target}^{\prime}}$, in $f^{\text{ASR-Enc}^{\prime}} (\cdot)$. 
Therefore, the prediction and joint networks are the same as the vanilla RNNT in~\ref{sssec:rnnt}. 
All networks with learnable parameters $\theta^{\text{TS-RNNT}} \triangleq [\theta^{\text{Spk-Enc}^{\prime}}, \theta^{\text{ASR-Enc}^{\prime}}, \theta^{\text{Pred}}, \theta^{\text{Joint}}]$ are jointly optimized by using RNNT loss.

When decoding with the TS-RNNT, we register the speaker embedding, $\bm{h}^{\text{target}^{\prime}}$, extracted from the enrollment with the speaker encoder in advance. 
Then, the acoustic features of the mixture signal, $f^{\text{FE}}(x_{t^{\prime}})$, is directly input to ASR encoder given also $\bm{h}^{\text{target}^{\prime}}$, and the ASR decoder yields the ASR results, $\hat{Y}$. 
Therefore, TS-RNNT can perform TS-ASR faster than the cascade system. 
Moreover, TS-RNNT does not need clean target-speaker's speech as reference for TSE module training, only the target-speaker's transcription for TS-RNNT training, which arguably can be easier to collect for real recordings. 

\vspace{-0.2cm}
\subsubsection{Streaming target-speaker ASR by TS-RNNT}
\vspace{-0.1cm}
In order to change TS-RNNT into a streaming model, 
we replace the ASR encoder of TS-RNNT with left-to-right encoding module as in~\cite{Sainath2021AnES,chen2021lcconformer}. 
Note that the speaker encoder module is not changed to streaming mode 
because it only needs to operate once to register the target-speaker's clues $\bm{h}^{\text{target}^{\prime}}$ in advance.
The performance of current TSE front-ends tend to degrade severely when operating in streaming mode. 
On the other-hand, ASR models can model well speech sequences with streaming models. 
By including the TSE functionality within the encoder of the RNNT, we expect to improve performance over cascade models for streaming mode. 
In addition, the tuning of the cascade system is more involving than that of the integrated system because we must separately tune the trade-offs of each module, 
i.e. TSE and ASR, between recognition performance and the latency. 
On the other hand, the integrated modeling, i.e. TS-RNNT, needs to tune just the ASR encoder to address the trade-off. 
Thus TS-RNNT would be easier to tune than cascade systems. 

\vspace{-0.2cm}
\section{Experiments}
\label{sec:result}
\vspace{-0.1cm}

\subsection{Experimental setups}
\vspace{-0.1cm}
\subsubsection{Data}
\label{ssec:data}
\vspace{-0.1cm}
We evaluated our proposal on speech recordings simulated using speech from the Corpus of Spontaneous Japanese (CSJ)~\cite{maekawa2000}, sampled at 16 kHz. It consists of two-speaker mixtures with background noise taken from the CHiME-4 corpus~\cite{chime4} at signal-to-noise ratio (SNR) between 0 and 20 dB.

We used almost the same data as in previous works~\cite{sato2021SE,Sato2021switch}. 
The details are shown in Table~\ref{tab:setup}. 
The total training data amounts for 800 hours of speech. 
The overlap ratio of both training and evaluation datasets was about 89\% on average. 
The speakers between training/development and evaluation datasets did not duplicate. 
We evaluated performance in terms of character error rate (CER) due to the ambiguity of Japanese word boundaries. 
We also measured inference speed in terms of real time factor (RTF: $decoding \ time / data \ time$). 
We measured the RTF using a Python implementation of the algorithm running on an Intel Xeon 2.40GHz CPU. 

\begin{table}[t]
\centering
\caption{CER as a function of the layer-$l$ where the fusion between the ASR encoder and speaker encoder output operates. The results are for offline TS-RNNTs. 
``All'' indicates that the fusion was applied to all layers of the encoder. 
}
 \vspace{-0.25cm}
\label{tab:comp}
\begin{tabular}{l|cccccc}
\hline
\multirow{2}{*}{\begin{tabular}[c]{@{}l@{}}Fusion in \\ layer-$l$\end{tabular}} & \multicolumn{6}{c}{CER {[}\%{]} on each SNR}                                                                                                 \\
                                                                              & 20dB                 & 15dB                 & 10dB                 & 5dB                  & \multicolumn{1}{c|}{0dB}  & Avg.                 \\ \hline
1                                                                             & \textbf{8.6}                  & \textbf{9.3}                  & \textbf{11.4}                 & \textbf{17.3}                 & \multicolumn{1}{c|}{\textbf{32.7}} & \textbf{15.8}                 \\
5                                                                             & 9.8                     & 10.4                     & 12.5                     & 18.0                      & \multicolumn{1}{c|}{33.1}     & 16.8                      \\
1-5                                                                           & 9.3                     & 9.8                     & 11.6                     & 17.6                     & \multicolumn{1}{c|}{32.9}     & 16.4                     \\
All                                                                          & 9.7 & 10.2 & 12.5 & 18.9 & \multicolumn{1}{c|}{35.1}     & \multicolumn{1}{l}{17.3} \\ \hline
\end{tabular}
\vspace{-0.125cm}
\end{table}

\begin{table}[t]
\centering
\caption{Comparisons of cascade and proposed TS-RNNT systems in terms of CERs and RTFs at each SNR. 
All models performed offline decoding. 
Note that RTFs exclude computation time of speaker encoder in TSE and TS-RNNT.
}
 \vspace{-0.25cm}
\label{tab:offline}
\hspace{-0.2cm}
\begin{tabular}{@{}l@{ }|@{ }cccccc|@{ }c@{}}
\hline
\multirow{2}{*}{System} & \multicolumn{6}{c|@{ }}{CER {[}\%{]} on each SNR {[}dB{]} }                & \multirow{2}{*}{RTF} \\
                        & 20 & 15 & 10 & 5  & \multicolumn{1}{c|}{0}  & Avg. &                      \\ \hline
RNNT                    & 76.0 & 75.7 & 75.4 & 75.7 & \multicolumn{1}{c|}{78.0} & 76.1     & \textbf{0.40}                 \\
TSE+RNNT                   &  \textbf{7.9}  &  \textbf{8.8}  & \textbf{11.1} & 18.3 & \multicolumn{1}{c|}{36.3} & 16.5 & 1.22                 \\ \hline
TS-RNNT                 & 8.6  & 9.3  & 11.4 & \textbf{17.3} & \multicolumn{1}{c|}{\textbf{32.7}} & \textbf{15.8} & \textbf{0.40}                 \\ \hline
\end{tabular}
\vspace{-0.525cm}
\end{table}

\begin{table*}[t]
\centering
\caption{Comparisons of cascade and proposed TS-RNNT systems. 
``\checkmark'' and ``\xmark'' indicate streaming or offline mode, respectively. Systems that have a ``\checkmark'' for the ``Streaming All'' column perform fully streaming inference. 
TS-RNNT does not use any TSE (``N/A''). }
 \vspace{-0.25cm}
 \label{tab:streaming}
\begin{tabular}{c|c|cc|c|ccc|cccccc}
\hline
\multirow{2}{*}{ID} & \multirow{2}{*}{\begin{tabular}[c]{@{}c@{}} ASR-Enc type\end{tabular}}  & \multicolumn{2}{c|}{\#frames of look-} & \multirow{2}{*}{System}   & \multicolumn{3}{c|}{Streaming}      & \multicolumn{6}{c}{CER {[}\%{]} on each SNR}                                                                       \\
 &                     &  \multicolumn{1}{c|}{back}     & ahead  &                                                                             & TSE & \multicolumn{1}{c|}{ASR} & All & 20dB          & 15dB          & 10dB          & 5dB           & \multicolumn{1}{c|}{0dB}           & Avg.          \\ \hline
1 &\multirow{3}{*}{Uni-Conformer} & \multicolumn{1}{c|}{\multirow{3}{*}{$\infty$}} & \multirow{3}{*}{3} & \multirow{2}{*}{TSE+RNNT}                                                                            & \xmark     & \multicolumn{1}{c|}{\checkmark}    & \xmark     & 11.3          & 12.5          & 16.0          & 24.6          & \multicolumn{1}{c|}{43.6}          & 21.6          \\
2 & & \multicolumn{1}{c|}{}                      & &                                                                                                        &  \checkmark   & \multicolumn{1}{c|}{\checkmark}    & \checkmark     &  16.9             & 18.9               & 24.1              & 36.2              & \multicolumn{1}{c|}{56.8}              & 30.6              \\ \cline{5-14} 
3 & & \multicolumn{1}{c|}{}                      & & TS-RNNT                                                                                          & N/A   & \multicolumn{1}{c|}{\checkmark}    & \checkmark    & \textbf{13.1} & \textbf{14.2} & \textbf{17.3} & \textbf{25.1} & \multicolumn{1}{c|}{\textbf{42.8}} & \textbf{22.5} \\ \hline \hline
4 & \multirow{3}{*}{LC-Conformer} & \multicolumn{1}{c|}{\multirow{3}{*}{$\infty$}} & \multirow{3}{*}{[0, 63]} & \multirow{2}{*}{TSE+RNNT}                                                                              & \xmark   & \multicolumn{1}{c|}{\checkmark}    & \xmark    & 10.0          & 11.0          & 14.2          & 22.5          & \multicolumn{1}{c|}{41.6}          & 19.9          \\
5 & & \multicolumn{1}{c|}{}                      & &                                                                                                       & \checkmark   & \multicolumn{1}{c|}{\checkmark}    & \checkmark     &   15.3            & 17.1              & 21.8              & 32.9              & \multicolumn{1}{c|}{53.5}              & 28.1              \\ \cline{5-14} 
6 & & \multicolumn{1}{c|}{}                      & & TS-RNNT                                                                                       & N/A   & \multicolumn{1}{c|}{\checkmark}    & \checkmark    & \textbf{11.4} & \textbf{12.4} & \textbf{15.0} & \textbf{21.9} & \multicolumn{1}{c|}{\textbf{38.6}} & \textbf{19.8} \\ \hline \hline
7 & \multirow{3}{*}{LC-Conformer} & \multicolumn{1}{c|}{\multirow{3}{*}{71}} & \multirow{3}{*}{[0, 63]} &\multirow{2}{*}{TSE+RNNT}                                                                            & \xmark   & \multicolumn{1}{c|}{\checkmark}    & \xmark    & 11.0          & 12.3          & 15.8          & 24.7          & \multicolumn{1}{c|}{44.2}          & 21.6          \\
8 &  & \multicolumn{1}{c|}{}                      & &                                                                                                       & \checkmark    & \multicolumn{1}{c|}{\checkmark}    & \checkmark    &  16.8             & 18.6              & 23.8              & 35.6              & \multicolumn{1}{c|}{56.3}              & 30.2               \\ \cline{5-14} 
9 & & \multicolumn{1}{c|}{}                      & &TS-RNNT                                                                                           & N/A   & \multicolumn{1}{c|}{\checkmark}    & \checkmark    & \textbf{11.9} & \textbf{13.0} & \textbf{16.0} & \textbf{23.8} & \multicolumn{1}{c|}{\textbf{41.2}} & \textbf{21.2} \\ \hline
\end{tabular}
\vspace{-0.4cm}
\end{table*}

\vspace{-0.2cm}
\subsubsection{System configuration of TSE module}
\label{ssec:sesystem}
\vspace{-0.1cm}

We adopted a time-domain SpeakerBeam structure as an TSE module~\cite{delcroix2020improving,sato2021SE}. 
It uses blocks of stacked 1D-convolution layers as proposed in ConvTasNet~\cite{luo2019conv}. 
The details of the implementation and training follows our prior work~\cite{sato2021SE}.
We also trained a streaming/causal TSE model by changing the convolution and global layer normalization (LN) to causal convolution and channel-wise LN, respectively. 
The algorithmic latency of the streaming TSE model is 1.25ms ($= 20 \text{samples} / 16k$), 
which is negligible for the latency of ASR decoding. 
All TSE models were trained with dataset (b) in Table~\ref{tab:setup}. 

\vspace{-0.2cm}
\subsubsection{System configuration of ASR module}
\label{ssec:sesystem}
\vspace{-0.1cm}

The input feature for ASR models was an 80-dimensional log Mel-filterbank. 
We used SpecAugment~\cite{specaugment} during training. 
In this paper, we adopted 3262 characters as the output tokens. 
The training and evaluation data were preprocessed following the Kaldi and ESPnet toolkits~\cite{povey2011kaldi,espnet}. 
The minibatch size was set to 64 in all experiments.
In this work, we adopted Conformer-based encoders as detailed below.

We investigated three version of (TS-) RNNT. 
First, we experimented with an offline system it consists of the same encoder architecture as Conformer (L)~\cite{anmol2020conformer} with a kernel size of 15. 
We used two-layer 2D-convolutional neural networks (CNNs) followed by 17 Conformer blocks,
with the stride sizes of both max-pooling layers at each layer set to 2. 
The prediction network had a 768-dimensional uni-directional LSTM layer followed by a 640-dimensional feed-forward layer. 
The speaker encoder for TS-RNNT had the same architecture as the ASR encoder while the number of blocks was reduced from 17 to 6 due to the memory usage during training.

We compare this system with two streaming systems that use a similar configuration as the offline system except that the encoder is replace with a causual Conformer (Uni-Conformer)~\cite{Sainath2021AnES} and latency-controlled Conformenr (LC-Conformer)~\cite{chen2021lcconformer}. 
All ASR encoders of the streaming systems replaced the depthwise convolution and batch normalization with the causal equivalent and LN, respectively, 
and were trained using an attention mask as in~\cite{chen2021lcconformer}.
The latency of Uni-Conformer with infinite history and few look-ahead frames, i.e. CNN module, was 30ms. 
The left and current chunk sizes of LC-Conformer were set to $\infty$/680ms and 600ms, respectively,
so the average latency was 330ms ($= 600\text{ms} / 2 + 30\text{ms}$). Here $\infty$ means that the encoder sees all past frames.

Offline Conformer model parameters were randomly initialized, 
while streaming Conformer parameters were initialized with trained offline Conformer parameters. 
``RNNT'' and ``TS-RNNT'' were trained with dataset (a) and (b) in Table~\ref{tab:setup}, respectively. 
All models were trained using RNNT loss by using the Adam optimizer with 25k warmup for a total of 100 epochs. 
For decoding, we used alignment-length synchronous decoding with beam size of 8~\cite{saon2020alsd}. 

\vspace{-0.1cm}
\subsection{Experimental results}
\vspace{-0.1cm}
\subsubsection{Ablation study} 
\vspace{-0.1cm}
First, we investigate the best position for the fusion of ASR encoder and speaker encoder outputs using offline TS-RNNTs.
Table~\ref{tab:comp} shows the CERs at each SNR. 
We can see that applying the speaker encoder output to just the first ASR encoder output, $l=1$, attained the best performance under all SNR conditions. 
Hereafter, we adopted $l=1$ for all TS-RNNT models. 

\vspace{-0.2cm}
\subsubsection{Cascade vs. integrated system using offline models} 
\vspace{-0.1cm}
Next, we compare the baseline cascade systems, TSE+RNNT, with the proposed integrated system, TS-RNNT for offline decoding. 
Table~\ref{tab:offline} shows the CERs and RTFs for each SNR condition. 
The RNNT baseline shows the CER obtained when recognizing the mixture without performing TSE using the RNNT back-end. 
As expected, this system performs poorly as it cannot identify the target speaker in a mixture.
By using the TSE model as front-end, TSE+RNNT could recognize the target-speaker's speech and achieve CER of 16.5\% on average. 
This result confirms the importance of the TSE module. 
However, the RTF of TSE+RNNT was much larger than that of RNNT decoding. 
We could reduce the RTF of the TSE+RNNT system by adopting a smaller TSE model such as ~\cite{wangvoicefilter}, 
but this would increase the CER. 
Regardless of how efficient this TSE front-end could be, it would inevitably increase the RTF. 

On the other hand, TS-RNNT achieves competitive or better CERs than the cascade system while equaling the RTF of RNNT. 
In particular, the CERs of TS-RNNT are better than those of the cascade system under severely noisy conditions, i.e. SNR 5 and 0 dB. 
This is probably because the TS-RNNT system is trained in E2E manner, and thus is not affected by processing artifacts that may limit performance of the TSE+RNNT system under severe conditions.
We could improve the performance of the TSE+RNNT cascade system by jointly training both modules or retraining the ASR back-end on processed speech~\cite{ShiZW2021jointmtasr}. 
However, the RTF of such a system would remain higher than our proposed TS-RNNT. 

\vspace{-0.2cm}
\subsubsection{Cascade vs. integrated system using streaming models} 
\vspace{-0.1cm}
Finally, we investigate the effectiveness of TS-RNNT for streaming models. 
Table~\ref{tab:streaming} shows the CERs under each SNR condition. 
``ASR-Enc type'' indicates the type of streaming Conformer, 
and ``\#frames'' is the number of look-back and look-ahead frames for the ASR encoders of the RNNT and TS-RNNT models. 
We investigated two types of LC-Conformer; LC-Conformer with infinite history context and LC-Conformer with the history context limited to 71 frames. 
A ``\checkmark'' and ``\xmark'' in the ``Streaming'' column indicates that the module is operating in streaming mode or offline mode, respectively. 
All systems are fully streaming systems except system ID 1, 4 and 7 that operate offline due to the TSE front-end. 
The SDR improvements in offline and streaming TSE models were 15.1dB and 11.1dB, respectively, which mirrors the tendency reported in prior separation studies~\cite{luo2019conv}. 

From the table, we observe that the proposed streaming TS-RNNT models (systems 3, 6 and 9) offer comparable performance to cascade TSE+RNNT with the offline TSE module (systems 1, 4 and 7), and greatly outperform streaming cascade systems, i.e., TSE+RNNT with the streaming TSE module (systems 2, 5 and 8) for all three encoder types. 
Comparing the performance of the offline TS-RNNT system of Table~\ref{tab:offline} and the best streaming system of Table~\ref{tab:streaming} (system 6), we observe a relative performance gap of about 20\%, which may be slightly worse than the gap observed for tasks with clean speech~\cite{moriya2022rnntadlm}. 
Closing this gap will be part of our future work. 

The LC-conformer with infinite lookback context (system 6) achieves the best overall performance with a latency of just 63 frames, i.e., 330 ms. 
Note that the performance of the LC-conformer degrades when we limit the lookback context (system 7, 8 and 9), which indicates that the future context is important, but so is the past context when processing overlapped speech. 

These results demonstrate the effectiveness of the integrated system for streaming decoding, as it avoids the performance degradation caused by using the streaming TSE module and achieves performance competitive with the offline TSE module while matching the inference speed of a vanilla RNNT.

\vspace{-0.2cm}
\section{Conclusion}
\label{ssec:conclusions}
\vspace{-0.1cm}
We have proposed an integrated modeling for a streaming TS-ASR, called TS-RNNT. 
The target-speaker embedding, which is extracted from the speaker encoder, is fused with the intermediate outputs of the RNNT encoder to allow direct recognition of the speech of a target speaker in a mixture. 
Our proposed system offers comparable performance to a cascade TSE+RNNT system in the offline setting, with significantly lower complexity. 
Indeed, our system kept the low complexity of a vanilla RNNT. 
Moreover, we can greatly outperform cascade systems in streaming mode.
The results of this work confirm the potential of TS-RNNT to achieve streaming ASR robust against interfering speakers.

\pagebreak

\bibliographystyle{IEEEtran}

\tiny{
\bibliography{mybib,refs}
}
\end{document}